\documentclass[11pt]{amsart}
 \usepackage[cp1251]{inputenc}
\usepackage[T2A]{fontenc}
\usepackage{latexsym}
\bigskip
\usepackage{graphicx} \usepackage{amsmath}
\newtheorem{lemma}{Lemma}[section]
\newtheorem{theorem}{Theorem}[section]
\newtheorem{definition}{Definition}[section]
\newtheorem{corollary}{Corollary}[section]
\newtheorem{example}{Example}[section]

\author{Dimiter Stoichkov Kovachev}
\address{
Department of computer science \\ South-West University "Neofit Rilski"\\
 Bulgaria, 2700 Blagoevgrad, P.O.79} \email{dim.kovach@gmail.com}

\title[On Some Classes of Functions and Hypercubes ]
{On Some Classes of Functions and Hypercubes }

\begin{document}
 
 \begin{abstract}
In this paper, some classes of discrete functions of $k$-valued logic are considered, that depend on sets of their variables in a particular way. Obtained results allow to "construct" these functions and to present them in their tabular, analytical or matrix form, that is, as hypercubes, and in particular Latin hypercubes. Results connected with identifying of variables of some classes of functions are obtained. \\

\end{abstract}

\maketitle

\section{  Introduction, Definitions and Notation}

 Let $E_k=\{ 0, \ 1,\ldots, \ k-1 \}, \ k\geq 2$. The set of all functions of $n$ variables of $k-$valued logic is denoted by $P_{n}^{k}$, where  $P_{n}^{k}=\{ f:E_k^n\longrightarrow E_k \}.$

A matrix of $m$ rows and $m$ columns is denoted by $||a_{ij}||_1^m$ and is called a $2-$dimensional matrix of order $m$.
By $||b_{i_{1}i_{2}...i_{n}}||_{1}^{k}$ we will denote the $n-$ dimensional matrix of order $k$, which is referred to as an  $n-$dimensional hypercube of order $k$ by some authors \cite{5}.
Each function $f(x_1, \ x_2,\ldots,\;x_n)\in P_{n}^{k}$ , by using the equality
\begin{equation} \label{1}
a_{i_{1}i_{2}...i_{n}}=f(x_{1}=i_{1}-1, \ x_{2}=i_{2}-1,\ldots,\;x_{n}=i_{n}-1),
\end{equation}
can be presented in the matrix form $||a_{i_{1}i_{2}...i_{n}}||_{1}^{k}$ as an $n-$ dimensional hypercube of order $k$,
based on the set $E_k$.

Latin squares and hypercubes have their applications \cite{5} in coding theory, error correcting codes, information security, decision making, statistics, cryptography, conflict-free access to parallel memory systems, experiment planning, tournament design, etc.

Each $n-$dimensional matrix $A=||a_{i_{1}i_{2}...i_{n}}||_{1}^{k}$ of order $k$
is called a Latin (Permutational) $n-$dimensional hypercube of order $k$, based on the
 set $E_k$, if for each $ s,~ s = 1, \ 2, \ldots , \ n, $ we have $\left|\displaystyle \bigcup_{j=1}^{k}\{a_{i_{1}...i_{s-1}\; {\bf j}
i_{s+1}...i_{n}} \}\right|  = |E_k|=k=$
$|\{a_{i_{1}...i_{s-1}\; {\bf 1}\; i_{s+1}...i_{n}}\}\cup
\{a_{i_{1}...i_{s-1}\; {\bf 2}\; i_{s+1}...i_{n}}\}\cup
...\cup \{a_{i_{1}...i_{s-1}\; {\bf k} \; i_{s+1}...i_{n}}\}|.$

Every function obtained from $f$  by replacing the variables of  $M$,  $M\subseteq X_f , 0\leq |M|\leq n,$ with constants is called a subfunction of $f$ with respect to $M$. The notation $g \prec f$ $(g\stackrel{M}\prec f)$
means that $g$ is a subfunction of  $f$  (with respect to $M$).

If $g \prec f$  ($g$ is a subfunction of $f$ ), then the matrix representation of $g$ is a hypercube which is a subhypercube of the hypercube of function $f$.

\cite{3} Range  of function $f(x_1,\; x_2,\ldots,\;x_n)$ is the number of different values which this function assumes, and by $Rng ( f )$ and $X_f =\{x_1,\; x_2,\ldots,\;x_n\}$ we denote the range and the set of variables of function $f$, respectively.

By $P_n^{k,q}, 1\leq q \leq k$ we denote the set of all functions belonging to $P_n^k$ and having range equal to $q$, that is, which assume exactly $q$ different values.

\begin{definition} \label {d1.1}
 \cite{3} If $M$ is a set of variables of the function $f$  and $G=\{ g: g\stackrel{\overline M}\prec f \}$ is the set of all subfunctions of $f$ with respect to $\overline {M} =X_{f} \setminus M$, then the set $ Spr(M, f)=\displaystyle \bigcup_{g\in G} \{ Rng(g )\}$ is called spectrum of the set $M$ for the function $f$.
\end{definition}

If $M=X_f$ , then $Spr(X_f, f ) =\{ Rng( f )\}$. For each function of one variable $g(x)\in P_1^k$, since $\{ x \}=X_g$, we have $Spr(\{ x \}, g ) = Spr(x, g ) =\{ Rng(g)\}$.

If for the function  $f$ we have $Spr(M, f)= \{q\}, 1\leq q \leq k$, this means that each subfunction $g$ of $f$ with respect to $M$ has range equal to $q$, that is, $Rng( g )=q$, and the matrix representation of $g$ is a hypercube that contains exactly $q$ different elements.

\begin{definition} \label {d1.2}
\cite{2} We say that $f(x_{1},\; x_{2},\ldots,\;
x_{n})$ is an $H-$function if for every variable $x_{i}$, $1\leq
i\leq n$, $n \geq 2$ and for every $n+1$ constants\\
$a_{1},\ldots,~a_{i-1},~a',~a'',~a_{i+1},\ldots,~a_{n}\in E_k$ with $a'\neq
a''$ we have
$$f(a_{1},\ldots,~a_{i-1},~a',~a_{i+1},\ldots,~a_{n})\neq
f(a_{1},\ldots,~a_{i-1},~a'',~a_{i+1},\ldots,~a_{n}).$$
\end{definition}

The matrix form of every $H-$function from $P_n^k$ is a Latin hypercube \cite{4}.

\section{  Main Results}

Let $a,\; b,\; c,\; a_i,\; b_i,\; i=1,\; 2,\ldots,\; n,$ be elements of the set $E_k$.

\begin{lemma}\label{le2.1}
 If the function $g(x_1,\; x_2,\ldots,\;x_n)\in P_n^{k,q} ,~ 1\leq q \leq k$ and

$f(x_1,\; x_2,\ldots,\;x_n)=(a.g + b)\; mod \; k$ where $(a,\; k)=1$, that is, $a$ and $k$ are coprime numbers, then $Rng( f )=Rng(g) =q$.
\end{lemma}

\begin{proof} From $g \in P_n^{k,q}$ , that is, $Rng(g)=q$, it follows that the function $g$ assumes $q$ different values.
 Let these values be $c_1,\; c_2,\ldots, c_q$, where $c_i\neq c_j$ when $i\neq j, \;i, j = 1,\; 2,\ldots,\; q$.
 Then $< ac_1+b,\;  ac_2+b,\ldots, ac_q+b >\; mod \; k$ are the values, assumed by function $f$.
 If these values are different with respect to $mod \; k$, it would follow that $Rng(g)=Rng( f )$.
Assume the contrary, that there exist $i \neq j$ such that $ac_i+b=ac_j+b \; mod \; k$.
After calculation we get $a(c_i - c_j)=0 \; mod \; k$, and since $(a, k)=1$, it follows that $c_i=c_j$,
which contradicts to the assumption that $c_i \neq c_j$.
The obtained contradiction is due to the assumption that among the values $ac_1+b,\;  ac_2+b,\ldots, ac_q+b $
there are values which are equal by $mod \; k$.
Therefore $f(x_1,\; x_2,\ldots,\;x_n)$ also assumes $q$ different values and hence $Rng( f )=Rng( g ) = q$.
\end{proof}

\begin{corollary} \label{co2.1}
Let  $f,\; g \in P_n^k$. If $f(x_1,\ldots,\;x_n)=(a.g + b)\; mod \; k$, $(a,\; k)=1$,
$M\subseteq X_f =X_g=\{ x_1,\ldots,\; x_n \}$, then $Spr(M, f ) = Spr(M, g )$.
\end{corollary}

We will say that a set $P$ is partitioned into the nonempty sets\\ $(P_1,\; P_2,\ldots,\; P_s),\;  s>1$, if:\\
 \centerline{$ 1)\;\; P_i \cap  P_j =\emptyset$, for $i \neq j,\; i,j \in \{ 1,\; 2,\ldots, s \};  \qquad 2)\;\; P = P_1 \cup P_2 \cup  \ldots \cup P_s$.}

Let the ordered $s-$tuple $S=( X_1,\; X_2,\ldots,\; X_s),\; 1<s \leq n$, be a partitioning of the set
$X_f =\{ x_1,~ x_2,\ldots,\;x_n \}$, and  vector  $\vec{q}=(q_1,\; q_2,\ldots, q_s),\; 1\leq q_i \leq k,\; i=1,\; 2,\ldots,\; s.$

\begin{definition} \label{d2.1}
 The function $f \in P_{n}^{k}$
is called $H(S, \vec{q})-$function if for each set of variables $X_i,\; i=1, \; 2,\ldots,\; s,$ we have $Spr(X_i, f)=\{ q_i \}$.
\end{definition}

If $Y= \{ x_{i_1}, \; x_{i_2}, \ldots, x_{i_r} \} $, for the sake of brevity, the function\\
 $h( x_{i_1}, \; x_{i_2}, \ldots, x_{i_r} ) $  is denoted by $h(Y)$.

\begin{theorem} \label{t2.1}
If the ordered $s-$tuple $S=( X_1,\; X_2,\ldots,\; X_s),\; 1<s \leq n$, is a partitioning of the set
$X_f =\{ x_1,\ldots,\;x_n \}$, vector  $\vec{q}=(q_1,\; q_2,\ldots, q_s),\; 1\leq q_i \leq k$, the functions
$f_i ( X_i)\in P_{|X_i|}^{k,q_i},\; (a_i,\; k)=1,\; i=1,\; 2,\ldots,\; s$, then the function
$f(x_1, \; x_2,\ldots,\;x_n)=[a_1 f_1 ( X_1) + a_2  f_2 ( X_2 ) +\ldots + a_s  f_s ( X_s) ]\; mod \; k$
is an $H(S, \vec{q})-$function.
\end{theorem}

\begin{proof} Let $X_i, \; 1 \leq i \leq s$, be an arbitrary set of variables and $g_i$ be an arbitrary subfunction of  $f$
with respect to $\overline{X_i}=X_f \setminus X_i$, that is, $g_i \stackrel{\overline {X_i}}\prec f$.
Since $g_i$ is obtained from $f$  by replacing all variables of $f$ from $\overline{X_i}$ with constants,
then $g_i(X_i)=[a_i f_i (X_i)+C_i] \; mod \; k$, where $C_i$ is a constant. From Lemma \ref{le2.1} it follows that
$Rng( g_i ) = Rng( f_i )=q_i$. Since subfunction $g_i$ was chosen arbitrarily, it follows that each subfunction
of $f$  with respect to $\overline{X_i}$  has a range equal to $q_i$ , and therefore $Spr(X_i, f )= \{ q_i \}$.
Because the set of variables $X_i$ was also chosen arbitrarily, it follows that for each $i,\; i=1,\; 2,\ldots,\; s$
we have $Spr(X_i, f ) = \{ q_i \}$, that is, the function $f(x_1,\ldots,\;x_n)$ is an $H(S, \vec{q})-$function.
\end{proof}
From  Theorem \ref{t2.1}, when $a_1= a_2= \cdots = a_s=1$, it follows:
\begin{corollary} \label{co2.2}
If the ordered $s-$tuple $S=( X_1,\; X_2,\ldots,\; X_s),\; 1<s \leq n$, is a partitioning of the set
$X_f =\{ x_1,\ldots,\;x_n \}$, vector $\vec{q}=(q_1,\; q_2,\ldots, q_s),\; 1\leq q_i \leq k$, the functions
$f_i ( X_i)\in P_{|X_i|}^{k,q_i},\; i=1,\; 2,\ldots,\; s$, then the function
$f(x_1, \; x_2,\ldots,\;x_n)=[f_1 ( X_1) + f_2 ( X_2 ) +\ldots + f_s ( X_s) ]\; mod \; k,$
is an $H(S, \vec{q})-$function.
\end{corollary}

\begin{example} \label{ex2.1}
"Construct" an $H(S, \vec{q})-$function of the set $P_3^3$, where\\
 $S=( X_1, X_2)$, $ \vec{q}=(3, 2), \;  X_1 = \{ x_1, x_3 \}, X_2 = \{ x_2 \}.$
\end{example}

Let $f_1 ( X_1) = f_1 ( x_1, x_3)$ and $f_2 ( X_2) = f_2 ( x_2)$ be arbitrary functions of the sets
$P_2^{3,3}$ and $P_1^{3,2}$, respectively, given in their tabular form in Table \ref{tab:1}.

\begin{table}[h]
\begin{center}
 \begin{tabular}{|c|c|c|c|c|c|c|c|c|c||c|c|c|}
  \hline
  {$x_1$} & {$ x_3$} & {$f_1$}    & {$x_1$} & {$x_3$} & {$f_1$}  & {$x_1$} & {$x_3$} & {$f_1$} & $Rng(f_1)$   &{$ x_2$} & {$f_2$} & $Rng(f_2)$\\   
\hline \hline
  0 & 0 & {\bf 2} & 1 & 0 & {\bf 1}  & 2 & 0 & {\bf 0} & 3 & 0 & {\bf 1} & 2 \\
  0 & 1 & {\bf 2} & 1 & 1 & {\bf 1}  & 2 & 1 & {\bf 2} &   & 1 & {\bf 0} &  \\
  0 & 2 & {\bf 0} & 1 & 2 & {\bf 1}  & 2 & 2 & {\bf 1} &   & 2 & {\bf 1} &  \\
  \hline
\end{tabular}
\end{center}
\caption {}
\label{tab:1}
\end{table}

According to Corollary \ref{co2.2}, the function $f (x_1,\; x_2,\; x_3)=[ f_1 ( x_1,\; x_3) + f_2 ( x_2)]\; mod \; 3$ is an $H(S, \vec{q})-$function of the set $P_3^3$ .
Consecutively we get: $f (0, 0, 0)=[ f_1 ( 0, 0) + f_2 ( 0)] \; mod \; 3 = [ 2 + 1]\; mod \; 3 = 0$, and so on, and results are systematized and entered in Table \ref{tab:2}. Except for in tabular form, according to equality (\ref{1}), the function $f (x_1, x_2, x_3)$ is also represented in matrix form.

\begin{table}[h]
\begin{center}
 \begin{tabular}{|c|c|c|c|c|c|c|c|c|c|c|c|c|c|c|c|c|}
  \hline
  {$x_1$} & {$x_2$} & {$x_3$} & {$a_{ijl}$} & {$f$} &   & {$x_1$} & {$x_2$} & {$x_3$} & {$a_{ijl}$} & {$f$} &   & {$x_1$} & {$x_2$} & {$x_3$} & {$a_{ijl}$} & {$f$} \\
  \hline \hline
  0 & 0 & 0 & $a_{111}$  &{\bf 0} &   & 1 & 0 & 0 & $a_{211}$  & {\bf 2} &   & 2 & 0 & 0 & $a_{311}$  & {\bf 1} \\
  0 & 0 & 1 & $a_{112}$  &{\bf 0} &   & 1 & 0 & 1 & $a_{212}$  & {\bf 2} &   & 2 & 0 & 1 & $a_{312}$  & {\bf 0} \\
  0 & 0 & 2 & $a_{113}$  &{\bf 1} &   & 1 & 0 & 2 & $a_{213}$  & {\bf 2} &   & 2 & 0 & 2 & $a_{313}$  & {\bf 2} \\
  \hline
  0 & 1 & 0 & $a_{121}$  &{\bf 2} &   & 1 & 1 & 0 & $a_{221}$  & {\bf 1} &   & 2 & 1 & 0 & $a_{321}$  & {\bf 0} \\
  0 & 1 & 1 & $a_{122}$  &{\bf 2 }&   & 1 & 1 & 1 & $a_{222}$  & {\bf 1} &   & 2 & 1 & 1 & $a_{322}$  & {\bf 2} \\
  0 & 1 & 2 & $a_{123}$  &{\bf 0} &   & 1 & 1 & 2 & $a_{223}$  & {\bf 1} &   & 2 & 1 & 2 & $a_{323}$  & {\bf 1} \\
  \hline
  0 & 2 & 0 & $a_{131}$  &{\bf 0 }&   & 1 & 2 & 0 & $a_{231}$  & {\bf 2} &   & 2 & 2 & 0 & $a_{331}$  & {\bf 1} \\
  0 & 2 & 1 & $a_{132}$  &{\bf 0 }&   & 1 & 2 & 1 & $a_{232}$  & {\bf 2} &   & 2 & 2 & 1 & $a_{332}$  & {\bf 0} \\
  0 & 2 & 2 & $a_{133}$  &{\bf 1 }&   & 1 & 2 & 2 & $a_{233}$  & {\bf 2} &   & 2 & 2 & 2 & $a_{333}$  & {\bf 2} \\
  \hline
\end{tabular}
\end{center}
\caption {}
\label{tab:2}
\end{table}

In the special case when $s=n$, $X_i =\{x_i\},~ i=1,~ 2,\ldots,~ n$, the function $f(x_1, \; x_2,\ldots,\;x_n)\in P_{n}^{k}$
is called $\vec{q} H-$function if for each variable $x_i$ we have  $Spr(x_i, f ) = \{ q_i \}$, where
$\vec{q}=(q_1,\; q_2,\ldots, q_n),\; 1\leq q_i \leq k,~i=1,~ 2,\ldots,~ n.$

From Theorem \ref{t2.1} we get:

\begin{corollary} \label{co2.3}
If the functions $f_i ( x_i)\in P_1^{k,q_i},~(a_i,\; k)=1,\;1\leq q_i \leq k,$\\
$i=1,~ 2,\ldots,~ n,$~vector $\vec{q}=(q_1,~q_2,\ldots, q_n)$, then the function\\
$f(x_1,~x_2,\ldots,x_n)=[a_1 f_1 ( x_1) + a_2  f_2 ( x_2 ) +\ldots + a_n  f_n ( x_n) ]\; mod \; k$
 is a $\vec{q} H-$function.
\end{corollary}

If $f$ is a $\vec{q} H-$function and in the hypercube corresponding to its matrix form we fix all indices, except the
 $i-$th index, by arbitrary values, then we obtain a one-dimensional matrix of order $k$,  which contains exactly
  $q_i, ~ i=1,~ 2,\ldots,~ n$, different elements.

In other words, if $||a_{i_{1}i_{2}...i_{n}}||_{1}^{k}$ is the matrix form of a $\vec{q} H-$function, where
$\vec{q}=(q_1,\; q_2,\ldots, q_n)$, then $ \left|\displaystyle \bigcup_{j=1}^{k}\{a_{i_{1}...i_{r-1}\; {\bf j}\;
i_{r+1}...i_{n}} \}\right|=q_r,~ r=1,~ 2,\ldots,~ n .$

Since each function $h \in P_1^{k,q_i}$ is of the form
$h=
\begin{pmatrix}
  0 & 1 & ... & k-1 \\
  b_1 & b_2 & ... & b_k
\end{pmatrix},~$ where
$b_i \in E_k,~ i=1,~ 2,\ldots,~ k$, then it can be written in the analytical form $y=h(x)$
by using an interpolating polynomial \cite{1} or in the following determinant form:\\
$$\left|
  \begin{array}{ccccccc}
    1 & x & x^2 & \ldots & x^{k-2} & x^{k-1} & y \\ [3pt]
    1 & 0 & 0 & \ldots & 0 & 0 & b_1 \\ [3pt]
       1 & 1 & 1 & \ldots & 1 & 1 & b_2 \\
    \vdots & \vdots & \vdots & \ddots & \vdots & \vdots & \vdots \\
    1 & k-1 & (k-1)^2 & \ldots & (k-1)^{k-2} & (k-1)^{k-1} & b_k \\
  \end{array}
\right| =0$$\\
Similarly to Example \ref{ex2.1}, using Corollary \ref{co2.3} we could "construct" an  $\vec{q} H-$ function which, in addition to  tabular and matrix form, could also be expressed in analytical form.

In the special case when $s=n$, $X_i =\{x_i\},~ i=1,~ 2,\ldots,~ n$, vector $\vec{q}=(q,\; q,\ldots, q)$, that is, $q_1= q_2=\ldots= q_n=q,
~ 1\leq q \leq k $, the function $f\in P_{n}^{k}$ is called an $H(q)-$function if for each variable $x_i$
 the following equality holds:  $Spr(x_i, f ) = \{ q \}$, $~i=1,~ 2,\ldots,~ n.$

From Theorem \ref{t2.1} we get:

\begin{corollary} \label{co2.4}
If the functions $f_i ( x_i)\in P_1^{k,q},~(a_i,\; k)=1,\;1\leq q \leq k,$\\
 $ i=1,~ 2,\ldots,~ n,~$ then the function
$f(x_1, \; x_2,\ldots,\;x_n)=[a_1 f_1 ( x_1) + a_2  f_2 ( x_2 ) +\ldots + a_n  f_n ( x_n) ]\; mod \; k$
 is a $H(q)-$function.
\end{corollary}

Similarly to Example \ref{ex2.1}, on the basis of Corollary \ref{co2.4} we can "construct" $H(q)-$functions.

\begin{theorem} \label{t2.2}
 The function  $f\in P_n^k$   is an $H(q)-$function if and only if each subfunction of $f$, depending on at least one variable, is an $H(q)-$function.
\end{theorem}

\begin{proof} (Necessity) Let the function $f\in P_n^k$ be an $H(q)-$function and $g$ be an arbitrary subfunction of $f$, for which $|X_g|\geq 1$. We will prove that $g$ is an $H(q)-$function.

Assume that $g$ is not an $H(q)-$function. Therefore, there exists a variable $x_r\in X_g$, such that $Spr(x_r, g ) \neq \{ q\}$, that is, there exists a subfunction $h$, $\{h\stackrel{X_{g}\setminus x_r}\prec g\} $, such that $Rng(h) \neq q. $

Since $h\stackrel{X_{g}\setminus x_r}\prec g $, $g \prec f$, it follows that $h\stackrel{X_{f}\setminus x_r}\prec f$.
From $Rng(h) \neq q$ and $h\stackrel{X_{f}\setminus x_r}\prec f$ it follows that $Spr(x_r, g ) \neq \{ q\}$ and $f$ is not an $H(q)-$function, a contradiction. Therefore, $g$ is an $H(q)-$function.

(Sufficiency) Let each subfunction of  $f$, depending on at least one variable, be an $H(q)-$function. We will prove that the function  $f$  is also an $H(q)-$function. Assume that  $f$  is not an $H(q)-$function, that is, there exists a variable $x_r\in X_f$ such that $Spr(x_r, f ) \neq \{  q\}$. Hence there exists a subfunction $g,~ g \stackrel {X_{f}\setminus x_r}\prec f$, $~g(x_r)\in P_1^k$, and since $\{ x_r \}=X_g, $
then $ Spr(x_r, g ) =\{ Rng(g) \} \neq \{ q \}$, that is, $g$ is not an $H(q)-$function, a contradiction. The obtained contradiction is due to the assumption that $f$  is not an $H(q)-$function.
\end{proof}

In the special case when $s=n$, $X_i =\{x_i\},~ i=1,~ 2,\ldots,~ n$, vector $\vec{q}=(k,\; k,\ldots, k)$, that is,
$q_1= q_2= \cdots = q_n=k$,
 the function $f \in P_{n}^{k}$ is called an $H(k)-$function or simply $H-$function if for every variable
 $x_i,~i=1,~ 2,\ldots,~ n$ we have  $Spr(x_i, f ) = \{ k \}$.

The reason for the above definition is the proved fact \cite{4} that matrix form of each $H-$function is a Latin hypercube and a function
$f \in P_{n}^{k}$ is an $H-$function if and only if for each variable $x_i,~i=1,~ 2,\ldots,~ n$ we have     $Spr(x_i, f ) = \{ k \}$.

Taking into account Theorem \ref{t2.1}, we obtain:

\begin{corollary} \label{co2.5}
If the functions $f_i ( x_i)\in P_1^{k,k}$, that is, they are bijective,\\
 $~(a_i,\; k)=1,~ i=1,~ 2,\ldots,~ n,~$ then the function
$f(x_1, \; x_2,\ldots,\;x_n)=[a_1 f_1 ( x_1) + a_2  f_2 ( x_2 ) +\ldots + a_n  f_n ( x_n) ]\; mod \; k$
 is an $H-$function, аnd its matrix form is an $n-$dimensional Latin hypercube of order $k$, based on the set $E_k$.
\end{corollary}

$H-$functions are special case of $H(S, \vec{q})-$function. All classes of functions, considered up to now, can also be viewed as a generalization of $H-$functions, аnd their matrix forms as a generalization of Latin hypercubes.

\begin{corollary} \label{co2.6}
If the functions $f_i ( x_i)$ are bijective, i.e. $f_i ( x_i)\in P_1^{k,k},~i=1,~ 2,\ldots,~ n$,
then the function
$f(x_1,\ldots,\;x_n)=[f_1 ( x_1) + f_2 ( x_2 ) +\ldots + f_n ( x_n) ]\; mod \; k$
 is an $H-$function.
\end{corollary}

From the fact that every function of the form $h(x)=ax+b$ is bijective and from Corollary \ref{co2.6} we get:

\begin{corollary} \label{co2.7}
\cite{4} If $(a_i,~ k)=1, ~i=1,~ 2,\ldots,~ n$, then the function\\
$f(x_1,\;x_2,\ldots,\;x_n)=[a_1 x_1 + a_2 x_2  +\ldots + a_n x_n ]\; mod \; k$
 is an $H-$function, and its matrix form is an  $n-$dimensional Latin hypercube of order $k$, based on the set $E_k$.
\end{corollary}

From Theorem \ref{t2.2} when $q=k$ we obtain:

\begin{corollary} \label{co2.8}
 \cite{7} A necessary and sufficient condition for the function  $f \in P_n^k$   to be an $H-$function is that each subfunction of $f$ depending on at least one variable to be an $H-$function.
\end{corollary}

 The function obtained from $f$  after replacing (identifying) variables\\ $x_{j_1},~ x_{j_2},\ldots,x_{j_t}$  by variable $z$ is denoted by  $f (x_{j_1}= x_{j_2}= \cdots =x_{j_t}=z)$.

\begin{theorem} \label{t2.3}
Let $f(x)\in P_1^{k,q},~g_i (x_i)=[ a_i f(x_i)+b_i ]~ mod~ k,~ (a_i, k)=1,~ i=1,~ 2,\ldots,~ n,$ and
$(a_{j_1}+a_{j_2}+\cdots +a_{j_t} ,~ k)=1,~ 1<t\leq n.$ When we identify variables $x_{j_1},~x_{j_2},~\ldots ,~x_{j_t}$ of the function
$$h(x_1,~ x_2,\ldots,~ x_n)=[ g_1 ( x_1)+ g_2 ( x_2 )+\ldots+ g_n ( x_n) ]~ mod~ k$$ with a new variable or with any of them, we obtain an $H(q)-$function belonging to the set $P_{n-t+1}^k$.
\end{theorem}

\begin{proof} From $f(x)\in P_1^{k,q},~g_i (x_i)=[ a_i f(x_i)+b_i ]~ mod~ k,~ (a_i, k)=1$ and Lemma \ref{le2.1} it follows that
$g_i (x_i)\in P_1^{k,q},~ i=1,~ 2,\ldots,~ n,$ According to Corollary \ref{co2.4} we can conclude that $h(x_1,~ x_2,\ldots,~ x_n)$
is an $H(q)-$function of $P_n^k$. Let $x_{j_1}=x_{j_2}= \cdots =x_{j_t}=x$. Then  $$g_{j_1} (x_{j_1})+ g_{j_2} (x_{j_2})+\cdots+
g_{j_t} (x_{j_t})$$ $$=\displaystyle \sum_{r=1}^t (a_{j_r} f (x)+b_{j_r})=(a_{j_1}+a_{j_2}+\cdots +a_{j_t}) f (x)+d,$$ where
$d=b_{j_1}+b_{j_2}+\cdots +b_{j_t}$. From $(a_{j_1}+a_{j_2}+\cdots +a_{j_t} ,~ k)=1,~f(x)\in P_1^{k,q}$
and Lemma \ref{le2.1} it follows that $[(a_{j_1}+a_{j_2}+\cdots +a_{j_t}) f (x)+d] ~ mod~ k$ is a function of $P_1^{k,q}$.
Applying again Corollary \ref{co2.4} to function  $h(x_{j_1}=x_{j_2}= \cdots =x_{j_t}=x)$, we complete the proof of the theorem.
\end{proof}

\begin{corollary} \label{co2.9}
Let $f(x)\in P_1^{k,q},~g_i (x_i)=[ a.f(x_i)+b_i ]~ mod~ k,~ (a, k)=1,~ i=1,~ 2,\ldots,~ n,$
$(t ,~ k)=1,~ 1<t\leq n.$ When identifying any $t$ variables of function
$h(x_1,~ x_2,\ldots,~ x_n)=[ g_1 ( x_1)+ g_2 ( x_2 )+\ldots+ g_n ( x_n) ]~ mod~ k$ with a new variable or with any of them,
we obtain an $H(q)-$function belonging to the set $P_{n-t+1}^k$.
\end{corollary}

When $q=k$, Theorem \ref{t2.3} and Corollary \ref{co2.9} refer to $H-$functions.

\begin{example} \label{ex2.2}
"Construct" an $H-$function $h(x_1,~ x_2,~ x_3)$ of the set $P_3^3$, such that $h(x_1=x_3=z) =h(z,~ x_2,~ z) $
and $h(z,~ x_2,~ z)$ be an $H-$function of $P_2^3$.
\end{example}
Let $f(x)\in P_1^{3,3}$ and $f(x)=
\begin{pmatrix}
  0 & 1 & 2  \\
  1 & 2 & 0
\end{pmatrix} =1+ {{5x-3x^2} \over 2}$, where the analytic expression is obtained by using an interpolating polynomial. Let
$h(x_1,~ x_2,~ x_3)=[ g_1 ( x_1)+ g_2 ( x_2 )+ g_3 ( x_3) ]~ mod~ 3$, where
$g_1 (x_1)=[2 f (x_1)+1]~ mod ~3=[2x_1]~ mod ~3,~  g_2 (x_2)=[ f (x_2)+2]~ mod~ 3 = [{{5x_2-3x_2^2} \over 2}]~ mod~ 3$,
$g_3 (x_3)= [2 f (x_3)+2]~ mod ~3 = [1+2x_3]~ mod~ 3.$ For the function h we get:\\
 $h(x_1,~ x_2,~ x_3)=[1+2x_1+ {{5x_2-3x_2^2} \over 2}+2x_3]~ mod~ 3.$ Then $$h_1(z, x_2)=h(z,~ x_2,~ z)=[1+ z +{{5x_2-3x_2^2} \over 2}]~ mod~ 3.$$

\begin{table}[h]
\begin{center}
\begin{tabular}{|c|c|c|c|c|c|c|c|c|c|c|c|c|c|c||c|c|c|c|}
  \hline
  $z$ & $x_2$ & $a_{ij}$ & $h_1$ &   & $z$ & $x_2$ & $a_{ij}$ & $h_1$ &   & $z$ & $x_2$ & $a_{ij}$ & $h_1$ &   &  ${{x_2\rightarrow}\over z\downarrow}$  & 0 & 1 & 2 \\
  \hline \hline
  0 & 0   & $a_{11}$ & {\bf 1}   &   & 1 & 0   & $a_{21}$ &{\bf 2}  &   & 2 & 0   & $a_{31}$ & {\bf 0}   &   & 0   &{\bf 1} &{\bf 2} &{\bf 0} \\
  0 & 1   & $a_{12}$ & {\bf 2}   &   & 1 & 1   & $a_{22}$ &{\bf 0}  &   & 2 & 1   & $a_{32}$ & {\bf 1}   &   & 1   &{\bf 2} &{\bf 0} &{\bf 1} \\
  0 & 2   & $a_{13}$ & {\bf 0}   &   & 1 & 2   & $a_{23}$ &{\bf 1} &   & 2 & 2   & $a_{33}$ & {\bf 2}   &   & 2   &{\bf 0} &{\bf 1} &{\bf 2} \\
  \hline
\end{tabular}
\end{center}
\caption {}
\label{tab:3}
\end{table}

In Table \ref{tab:3}, the function $h_1(z, x_2)$ is given  in both tabular and matrix form.

\end{document}